\documentclass[12pt]{article}
\usepackage{graphicx}
\usepackage{a4p}
\usepackage{color}
\usepackage{amsmath}
\usepackage{hyperref}  
\definecolor{darkblue}{rgb}{0,0,.5}
\hypersetup{
linktocpage=true,
    bookmarks=true,         
breaklinks=true,
    frenchlinks=false, %
    unicode=false,          
    pdftoolbar=true,        
    pdfmenubar=true,        
    pdffitwindow=false,     
    pdfstartview={FitH},    
    pdfauthor={J.Fleischer, T.Riemann},     
    pdfcreator={JF+TR},   
    pdfproducer={JF+TR}, 
    pdfkeywords={}, 
    pdfnewwindow=true,      
    colorlinks=false,       
    linkcolor=red,          
    citecolor=green,        
    filecolor=magenta,      
    urlcolor=cyan           
}
\newcommand{\bea}{\begin{eqnarray}} 
\newcommand{\eea}{ \end{eqnarray}} 
\newcommand{\ba}{\begin{eqnarray*}}
\newcommand{\ea}{  \end{eqnarray*}}
\newcommand{\nn}{\nonumber}

\newcommand{\beq}{\begin{equation*} }
\newcommand{\eeq}{\end{equation*}  }

\newcommand{\bqa}{\begin{eqnarray*} }
\newcommand{\eqa}{\end{eqnarray*}}
\newcommand{\nl}{\nonumber \\ \nonumber}

\def \eps {\epsilon}

\begin{document}
{%
\texttt{\small%
\begin{flushleft}
DESY 11-225%
\\
BI-TP 2011/48%
\\
SFB/CPP-11-76%
\\[0mm]
LPN 11-72\\[8mm] 
\end{flushleft}}

\bigskip

\begin{center}
{\LARGE \bf
One-loop tensor Feynman integral reduction with signed minors}
\\
\vspace{1.0cm}

\renewcommand{\thefootnote}{\fnsymbol{footnote}}
{\Large J. Fleischer~\footnote{E-mail:~fleischer@physik.uni-bielefeld.de}${}^{a}$, 
   T. Riemann~\footnote{E-mail:~Tord.Riemann@desy.de}${}^{b}$ 
 ~~and~~
        V. Yundin~\footnote{E-mail:~yundin@nbi.dk}${}^{c}$        }
\\[1cm]
\end{center}

{\noindent
{${}^{a}$~Fakult\"at f\"ur Physik, Universit\"at Bielefeld, Universit\"atsstr. 25,  33615
Bielefeld, Germany }
\\\noindent
{${}^{b}$~Deutsches Elektronen-Synchrotron, DESY, Platanenallee
  6, 15738 Zeuthen, Germany}
\\\noindent
{${}^{c}$~Niels Bohr International Academy and Discovery Center, Niels Bohr
Institute, University of Copenhagen, Blegdamsvej 17, DK-2100, Copenhagen,
Denmark
}
}

%
%

\begin{abstract}
We present an algebraic approach to one-loop tensor integral reduction.
The integrals are presented in terms of scalar one- to four-point functions.
The reduction is worked out explicitly until five-point functions of rank five.
The numerical C++ package PJFry evaluates tensor coefficients in terms of a basis of scalar integrals, which is provided by an external library, e.g. QCDLoop.
We shortly describe installation and use of PJFry.
Examples for numerical results are shown, including a special treatment for small or vanishing inverse four-point Gram determinants.
An extremely efficient application of the formalism is the immediate evaluation of complete contractions of the tensor integrals with external momenta. This leads to the problem of evaluating sums over products of signed minors with scalar products of chords. Chords are differences of external momenta. These sums may be evaluated analytically in a systematic way. The final expressions for the numerical evaluation are then compact combinations of the contributing basic scalar functions.  
\end{abstract}

\section{Tensor reductions}
The reduction of tensorial Feynman integrals to scalar Feynman integrals is an old technique.
A systematic approach has been developed for one- to four-point functions in \cite{Passarino:1978jh}.
The simple case of a self-energy vector
\bea
\label{definition}
 I_2^{\mu} &=&  \int \frac{d^d k}{i\pi^{d/2}}~~\frac{k^{\mu}}{[k^2-M_1^2]~[(k+p)^2-M_2^2]}
\\\nonumber
&=& p^{\mu} B_1
\eea
is evaluated in terms of scalar functions $A_0, B_0$ as follows:
 \bea
 p_{\mu} I_2^{\mu} &=& p^2 { B_1(p, M_1,M_2)}
\nl
&=&
 \int \frac{d^d k}{i\pi^{d/2}}~~\frac{pk}{[k^2-M_1^2]~[(k+p)^2-M_2^2]} ~~\equiv~~ \int \frac{d^d k}{i\pi^{d/2}}~~\frac{pk}{D_1~D_2}
\\ &=&
 \frac{1}{2} \int \frac{d^d k}{i\pi^{d/2}}~~\left[
\frac{ D_2 - (p^2-M_2^2-M_1^2)  -D_1   }{D_1~D_2}
 \right] ,
\\
 { B_1 (p, M_1,M_2})&=& \frac{1}{2p^2}\left[  { A_0(M_1)} -  {  A_0(M_2)} - \frac{1}{2}(p^2-M_2^2-M_1^2) { B_0(p, M_1,M_2)}\right] .
\eea
This works fine for many situations, but for higher-point functions and for specific kinematical situations other approaches are more useful.
We advocate here Davydychev's approach \cite{Davydychev:1991va} and represent $n$-point tensors by $n$-point scalars in higher dimensions and with higher indices (powers of the propagators $D_i=1/[(k-q_i)^2 -m_i^2]$).
An example is:
\begin{eqnarray}
\label{tensor3}
I_{5}^{\mu  \nu  \lambda}
&=& \int
\frac{d^{d}  k}{{i\pi}^{d/2}}
   \frac{ k^{\mu}   k^{\nu}    k^{\lambda} }{D_1 \cdots D_5}
\\ \nonumber
 &=& -~ \sum_{i,j,k=1}^{4}
 { q_i^{\mu}  q_j^{\nu}  q_k^{\lambda} }
 n_{ijk}
 { I_{5,ijk}^{[d+]^3}}
+
\frac{1}{2} \sum_{i=1}^{n-1}
  {
(  g^{\mu \nu}    q_i^{\lambda}  + g^{\mu \lambda}    q_i^{\nu}  +
      g^{\nu \lambda}    q_i^{\mu}   ) }
{  I_{5,i}^{[d+]^2} }  ,
\end{eqnarray}
with $ n_{ijk} = (1+\delta_{ij} ) (1+\delta_{ik} + \delta_{jk}) $.
The $I_{5,ijk}^{[d+]^L}$  is defined here in $D=d-2\eps+2L$ dimensions  and propagators of lines $i,j,k$ have a power raised by one unit.  
In a next step Tarasov's dimensional recurrences \cite{Tarasov:1996br,Fleischer:1999hq} may be applied in order to diminish the dimensions of the scalar integrals:\footnote{The explanation of notations may be found in any of the articles \cite{Diakonidis:2008ij,Tarasov:1996br,Fleischer:1999hq,Fleischer:2010sq}.For Gram determinants and signed minors see \ref{app-det}.} 
\begin{equation}
\label{eq:RR1}
 \nu_j    \bigl( {\bf j^+} I_5^{{ [d+]}} \bigr)
=
\frac{1}{{  \left( \right)_5}}
\left[  - {{j \choose 0}_5} +\sum_{k=1}^{5} {j \choose k}_5
   {{\bf k^-} }\right]   {I_5 } ,
\end{equation}
and
\begin{equation} 
\label{eq:RR2}  
  (d-\sum_{i=1}^{5}\nu_i+1)      I_5^{{ [d+]}}
=
\frac{1}{ {{\left( \right)_5}}}
  \left[ {{0 \choose 0}_5}
 - \sum_{k=1}^5 {0 \choose k}_5 {{\bf k^-}} \right]   I_5,
\end{equation}
%
where the operators   ${\bf  i^{\pm}, j^{\pm}, k^{\pm} }$ act by shifting
the indices   {$\nu_i, \nu_j, \nu_k$} by $\pm 1$.

After dedicated simplifications \cite{Diakonidis:2008ij,Fleischer:2010sq}, the tensor of rank $R=3$  may be written as follows:
\begin{eqnarray}
I_{5}^{\mu\, \nu\, \lambda} &=& \sum_{i,j,k=1}^{4} \, q_i^{\mu}\, q_j^{\nu} \, q_k^{\lambda}
E_{ijk}+\sum_{k=1}^4 g^{[\mu \nu} q_k^{\lambda]} E_{00k},
\label{Exyz0}
\end{eqnarray}
with the tensor coefficients
\bea
\label{Exyz1}
E_{00j} 
&=&
 \sum_{s=1}^5 \frac{1}{ { {0\choose 0}_5} }   \left[\frac{1}{2} {0s\choose 0j}_5
 { I_4^{[d+],s} }
- \frac{d-1}{3} {s\choose j}_5
{ I_4^{[d+]^2,s} } \right]  ,
\\
E_{ijk} 
&=&
-  \sum_{s=1}^5\frac{1}{{{0\choose 0}_5}} \left\{ \left[{0j\choose sk}_5
{ I_{4,i}^{[d+]^2,s}}
+
(i \leftrightarrow j)\right]+{0s\choose 0k}_5 {\nu}_{ij}
{I_{4,ij}^{[d+]^2,s}} \right\} .
\label{Exyz2}
\eea
At this stage, we have reached:
\begin{itemize}
 \item All scalar 5-point integrals are eliminated.
\item No inverse Gram determinants $()_5$ have appeared.
 \item Scalar 4-point integrals in higher dimensions and/or with higher indices are present yet, e.g. {$ I_{4,ij}^{[d+]^2,s}$}.
\item Inverse Cayley determinants ${0\choose 0}_5$ have appeared.
\end{itemize}
The integrals $ I_{4,ij}^{[d+]^2,s}$ arise from 5-point integrals by shrinking line $s$.

By further, nontrivial manipulations \cite{Fleischer:2010sq} we have avoided shifted indices and isolated the inverse Gram determinants $()_4$ in the higher-dimensional scalar 4-point functions; see e.g. in \eqref{Exyz2}: 
\bea
{I_{4,i}^{[d+],s}} 
&=&
\frac{1}{{0s\choose 0s}_5}\left[-{0s\choose is}_5 (d-3) {I_4^{[d+],s}}
+\sum_{t=1}^5 {0st\choose 0si}_5{ I_3^{st}}\right]  ,
\label{I4id+2}
\\
{\nu}_{ij}{ I_{4,ij}^{[d+]^2}}&=& ~~\frac{{0\choose i}_4 }{{0\choose 0}_4 }\frac{{0\choose j}_4 }{{0\choose 0}_4 }
(d-2)(d-1){I_4^{[d+]^2}} +\frac{{0i\choose 0j}_4 }{{0\choose 0}_4 }{I_{4}^{[d+]}}
\nn \\
&&-\frac{{0\choose j}_4}{{0\choose 0}_4}\frac{d-2}{{0\choose 0}_4}\sum_{t=1}^4 {0t\choose 0i}_4{I_3^{[d+],t}} +
\frac{1}{{0\choose 0}_4}\sum_{t=1}^4 {0t\choose 0j}_4{I_{3,i}^{[d+],t}} .
\label{want1}
\eea
These equations contain yet the generic 4-point and (partly indexed) 3-point functions in higher dimensions,
${I_4^{[d+],s}},~~{I_{3,i}^{[d+],t}} $, etc.
{Several strategies are now possible:}
\begin{itemize}
 \item Evaluate them {analytically} in $d+2l-2\eps$  dimensions -- if you may do that.
 \item Evaluate them {numerically} in $d+2l-2\eps$  dimensions.
 \item {Reduce} them further by recurrences $\rightarrow$ apply \eqref{eq:RR2} -- but buy some towers of {$1/()_4$} at this stage.
 \item Make a {small Gram determinant expansion} $\rightarrow$ apply \eqref{eq:RR2} the other way round
\end{itemize}
The last two items are done here.

{For non-small 4-point Gram determinants}, the direct, iterative  use of \eqref{eq:RR2} yields
\begin{eqnarray}
\label{A401}
I_{4}^{[d+]^l}&=&\left[\frac{{0\choose 0}_4}{\left(  \right)_4}I_{4}^{[d+]^{l-1}}-
\sum_{t=1}^{4} \frac{{t\choose 0}_4} {\left(  \right)_4} I_{3}^{[d+]^{l-1},t}  \right]
\frac{1}{d+2 l-5} ,
\\
 I_{3  }^{[d+]^l,t}&=& \left[ \frac{{0t\choose 0t}_4}{{t\choose t}_4}  I_{3  }^{[d+]^{l-1},t}-
\sum_{u=1, u \ne t}^4 \frac{ {ut\choose 0t}_4}{{t\choose t}_4} I_{2  }^{[d+]^{l-1},tu} \right]\frac{1}{d+2 l-4},
\label{A301}
\end{eqnarray}
and we are done.
This works fine if $()_4$ is not small [and also the ${t\choose t}_4$].

For small  $()_4$, one may apply the recurrence relation the other way round and get:
 \bea
\label{eqarb4}
 I_n^{D} &=&
  \sum_k^n \frac{{0 \choose k}_n}{ {0 \choose 0}_n}   I_{n-1}^{D,k}
\nl &&
-~ {\frac{()_n}{ {0 \choose 0}_n }} [(D+1)-\sum_i^n\nu_i]
\nonumber \\ && \times~
\left[
  \sum_k^n \frac{{0 \choose k}_n}{ {0 \choose 0}_n}   I_{n-1}^{D+2,k} - {\frac{()_n}{ {0 \choose 0}_n }} [(D+3)-\sum_i^n\nu_i] I_n^{D+4}
\right] + \cdots
\eea
This works in fact fine, especially when improved by Pade approximations. For explicit examples see \cite{Fleischer:2010sq} and sec. \ref{pjfry} on the PJFry package here.
The approach has been independently implemented and used by Reina et al. \cite{Reina:2011mb}.


\section{PJFry\label{pjfry}}
The goal of the C++ package PJFry is a stable and fast open-source implementation of tensor reduction, suitable for any physically relevant kinematics.\footnote{The presentation relies partly on: V. Yundin, One loop tensor reduction program PJFRY, talk at meeting of SFB/TR9, 15 Nov. 2011, Aachen, Germany.}
It performs the reduction of 5-point 1-loop tensor integrals up to rank 5. The
4- and 3-point tensor integrals are obtained as a by-product.
Main features are:
\begin{itemize}
 \item Any combination of (real) internal or external masses
\item Automatic selection of optimal formula for each coefficient
\item Leading $()_5$ are eliminated in the reduction
\item Small $()_4$ are avoided using asymptotic expansions where appropriate
\item Cache system for tensor coefficients and signed minors
\item Interfaces for C, C++, FORTRAN and Mathematica
\item Uses QCDLoop \cite{Ellis:2007qk,vanOldenborgh:1990yc} or OneLOop \cite{vanHameren:2010cp} for 4-dim scalar integrals
\item Available from the project webpage \url{https://github.com/Vayu/PJFry/} \cite{pjfry:2011aa,Fleischer:2010sq}
\end{itemize}
The installation of PJFry may be performed following the instructions given at the project webpage. 
\\
The project subdirectories are\\
./src       - the library source code\\
./mlink     - the MathLink interface\\
./examples  - the FORTRAN examples of library use, built with make check

\bigskip

A build on Unix/Linux and similar systems is done in a standard way by sequential performing
   ./configure, make, make install.
See the INSTALL file for a detailed description of the ./configure options.

The PJFry is used as one option of the GoSam package \cite{Cullen:2011ac}.

The functions for tensor coeffcients for up to rank-5 pentagon integrals are declared in the Mathematica interface:
\begin{verbatim}
In:= Names["PJFry`*"]Names[PJFry]

{A0v0, B0v0, B0v1, B0v2, C0v0, C0v1, C0v2, C0v3, \
D0v0, D0v1, D0v2, D0v3, D0v4, E0v0, E0v1, E0v2, \
E0v3, E0v4, E0v5, GetMu2, SetMu2}

Out= {A0v0, B0v0, B0v1, B0v2, C0v0, C0v1, C0v2, C0v3, \
D0v0, D0v1, D0v2, D0v3, D0v4, E0v0, E0v1, E0v2, \
E0v3, E0v4, E0v5, GetMu2, SetMu2}
\end{verbatim}
The syntax is very close to that of e.g. LoopTools/FF:
\begin{verbatim}
E0v3[i,j,k,p1s,p2s,p3s,p4s,p5s,s12,s23,s34,s45,s15,m1s,m2s,m3s,m4s,m5s,ep=0]
\end{verbatim}
where: 
\\
$i,j,k$ are indices of the tensor coefficient $(0 < i \leq  j \leq  k < n)$, 
\\
$p1s,p2s,...$ are squared external masses $p_i^2$, 
\\
$s12,s23,...$ are Mandelstam invariants $(p_i + p_j)^2$,
\\
$m1s,m2s,...$ are squared internal masses $m_i^2$,
\\
$ ep=0,-1,-2$ selects the coefficient of the $\eps$-expansion.

The average evaluation time per phase-space point on a 2 GHz Core 2 
laptop for the evaluation of all 81 rank 5 tensor form-factors amounts to 2 ms.

Numerical examples are shown in figures \ref{fig-big} and \ref{fig-small} for a
rank $R=4$ tensor coefficient in a region, where one of the 4-point sub-Gram determinants becomes small

\bigskip

$E_{3333}(0,0,-6{\times}10^4(x+1),0,0,%
             10^4,-3.5{\times}10^4,2{\times}10^4,-4{\times}10^4,1.5{\times}10^4,%
             0,6550,0,0,8315)$

\bigskip

When $x=0$, the 4-point Gram determinant vanishes; see \ref{app-det} for details.

   \begin{figure}[t]
        \centering
	\includegraphics[angle=0,width=.8\textwidth]{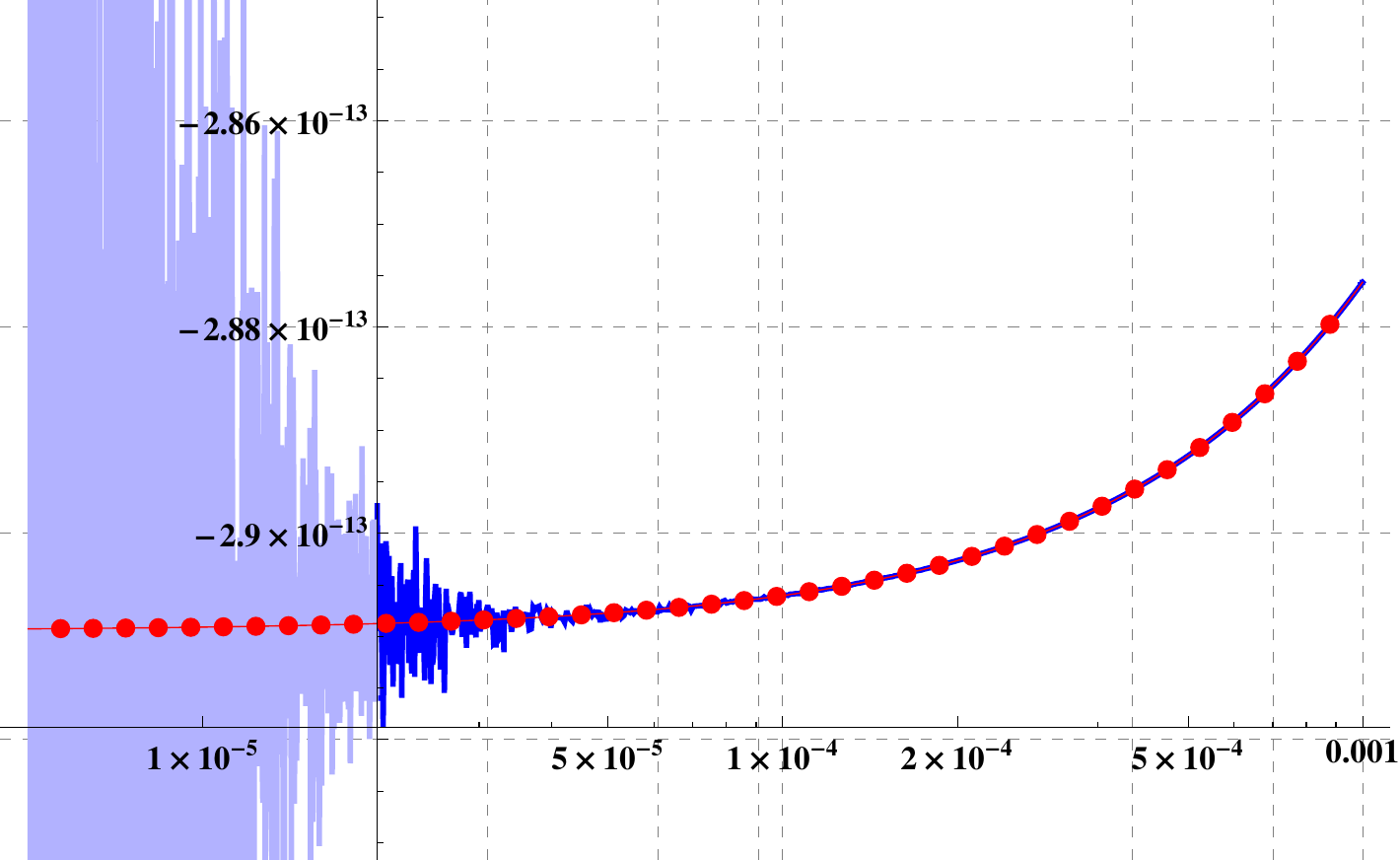}
\caption{Absolute accuracy of $E_{3333}$ near  the region of small Gram determinant.
Blue curve: conventional Passarino-Veltman reduction, red curve: PJFry.\label{fig-big}}
    \end{figure}

   \begin{figure}[t]
        \centering
        \includegraphics[angle=0,width=.9\textwidth]
{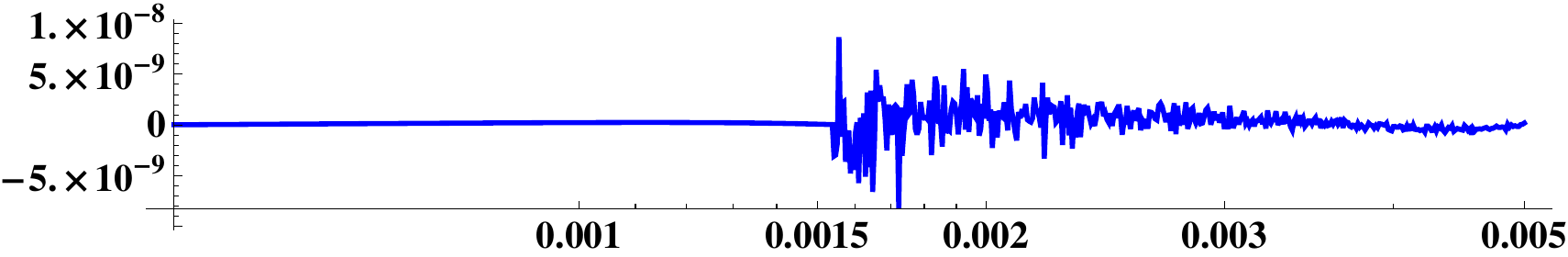}
\caption{Relative accuracy of $E_{3333}$ near the region of small Gram determinant. At $x \sim 0.0015$ PJFry switched to the asymtotic expansion.
\label{fig-small}}
    \end{figure}

\section{Contractions of tensor integrals with external momenta}
The contents of this section is worked out in detail in \cite{Fleischer:2011nt}.
A starting point are representations of tensors in terms of 
 basis functions $I_n^D$, which are independent of the tensor indices \{$i,j,k,...$\}.
Then, 
one may use {contractions with external momenta} in order to perform all the sums over  \{$i,j,k,...$\}.
This will evidently lead to a {significant simplification and shortening} of calculations.

To demonstrate the idea, let us look at the most involved 5-point tensor studied:
\bea\label{compl5a}
I_{5}^{\mu \nu \lambda \rho \sigma} &=&  
\sum_{s=1}^5 \Biggl[
\sum_{i,j,k,l,m=1}^{5}  q_i^{\mu} q_j^{\nu}  q_k^{\lambda} q_l^{\rho} q_m^{\sigma} E_{ijklm}^s
+
\sum_{i,j,k=1}^5 g^{[\mu \nu} q_i^{\lambda} q_j^{\rho} q_k^{\sigma]} E_{00ijk}^s
+
\sum_{i=1}^5
g^{[\mu \nu} g^{\lambda \rho} q_i^{\sigma]} E_{0000i}^s \Biggr] .
\label{compl5}
\nonumber \\
\eea
In the approach avoiding inverse Gram  determinants $1/()_5$, we have
\bea
E_{ijklm}^s &=&
-\frac{1}{{0\choose 0}_5}  \Biggl\{ \left[
{0l\choose sm}_5 {n}_{ijk}I_{4,ijk}^{[d+]^4,s}+(i \leftrightarrow l)+(j \leftrightarrow l)+
(k \leftrightarrow l)\right]
+{0s\choose 0m}_5{n}_{ijkl} I_{4,ijkl}^{[d+]^4,s} \Biggr\} 
\nonumber \\
\label{Ewxyz5}
\eea
and simpler expressions.
Now, in a next step, one may avoid the appearance of inverse sub-Gram determinants $()_4$ and eliminate the higher indices of propagators \{$i,j,k,\ldots$\}.
Then, the complete dependence on the indices $i$   of the tensor coefficients is contained in the pre-factors with signed minors:
\emph{the tensor indices {decouple} from the integrals}. 

As an example, we reproduce the 4-point part of $I_{4,ijkl}^{[d+]^4}$:
\bea
&&{n}_{ijkl} I_{4,ijkl}^{[d+]^4}=
\frac{{0\choose i}}{{0\choose 0}}
\frac{{0\choose j}}{{0\choose 0}}\frac{{0\choose k}}{{0\choose 0}}\frac{{0\choose l}}{{0\choose 0}}
d(d+1)(d+2)(d+3)I_4^{[d+]^4} \nn \\
&&+~\frac{{0i\choose 0j}{0\choose k}{0\choose l}+{0i\choose 0k}{0\choose j}{0\choose l}+{0j\choose 0k}{0\choose i}{0\choose l}+
{0i\choose 0l}{0\choose j}{0\choose k}+{0j\choose 0l}{0\choose i}{0\choose k}+{0k\choose 0l}{0\choose i}{0\choose j}}
{{0\choose 0}^3}  d(d+1)I_4^{[d+]^3}\nn \\
&&+~\frac{{0i\choose 0l}{0j\choose 0k}+{0j\choose 0l}{0i\choose 0k}+{0k\choose 0l}{0i\choose 0j}}
{{0\choose 0}^2}I_4^{[d+]^2} 
~~+~ \cdots
\label{fulld4}
\eea
In \eqref{fulld4}, the 3-point terms are not reproduced, and
one has to understand the 4-point integrals to carry the corresponding index $s$ and the signed minors are ${0\choose k}\to {0s\choose ks}_5$ etc.

A chord is the momentum shift of an internal line due to external momenta,  $D_i=(k-q_i)^2-m_i^2+i\epsilon$, and
$q_i =−(p_1 + p_2+· · ·+ p_i)$, with $q_n = 0$.
{The tensor $5$-point integral of rank $R=1$}  
yields, when contracted with a chord,
\begin{eqnarray}\label{i5vc1}
q_{a \mu} I_{5}^{\mu}&=& - \frac{1 }{{0\choose 0}_5} 
\sum_{s=1}^{5}
\left[
\sum_{i=1}^{4} (q_{a} \cdot q_i) {0i\choose 0s}_5 \right]
 I_{4}^{s}.
\label{i5vc2}
\end{eqnarray}
In fact, the {sum over $i$ may be performed explicitly}:
\bea 
\label{eq-wa-83a}
{\Sigma}^{1,s}_{a} 
&\equiv&
\sum_{i=1}^{4}(q_a \cdot q_i)  {0s\choose 0i}_5 
~=~ +\frac{1}{2} \left\{
{s\choose 0}_5\left(Y_{a5}-Y_{55} \right)+
{0\choose 0}_5\left({\delta}_{as}-{\delta}_{5s}\right) \right\}, 
\eea
and
we get immediately the desired compact result for the contraction of chords (or external momenta) and tensor integrals:
\bea
q_{a \mu}  I_{5}^{\mu}= 
-~\frac{1 }{{0\choose 0}_5}\sum_{s=1}^{5}{\Sigma}_{a}^{1,s}~~I_{4}^{s}.
\eea
Similarly, the tensor $5$-point integral of rank $R=2$ may be  treated:
\begin{eqnarray}
I_{5}^{\mu\, \nu\,}= \sum_{i,j=1}^{4} \, q_i^{\mu}\, q_j^{\nu} E_{ij} +
g^{\mu \nu}  E_{00} ,
\label{final2}
\end{eqnarray}
It has the following tensor coefficients free of $1/()_5$:
\bea\label{E00}
E_{00}
&=&
 - \sum_{s=1}^5    \frac{1}{2} \frac{1}{{0\choose 0}_5} {s\choose 0}_5 I_4^{[d+],s},
\\
E_{ij} 
\label{Exy}
&=&
\sum_{s=1}^5   \frac{1}{{0\choose 0}_5}   \left[{0i\choose sj}_5 I_4^{[d+],s}+
{0s\choose 0j}_5 I_{4,i}^{[d+],s} \right].
\label{Eij}
\eea
Equation \eqref{final2}  yields for the contractions with chords:
\bea\label{qqi2}
q_{a \mu} q_{b \nu} I_{5}^{\mu\, \nu\,}= \sum_{i,j=1}^4 (q_a \cdot q_i) (q_b \cdot q_j) E_{ij}+
(q_a \cdot q_b) E_{00} ,
\eea
 and finally \eqref{qqi2} simply reads
\bea
q_{a \mu} q_{b \nu} I_{5}^{\mu\, \nu\,}
&=&\frac{1}{4}
\sum_{s=1}^{5}\Biggl\{
\frac{{s\choose 0}_5}{{0s\choose 0s}_5}({\delta}_{ab}{\delta}_{as}+{\delta}_{5s}) 
+\frac{{s\choose s}_5}{{0s\choose 0s}_5} \Bigl[ 
\left({\delta}_{as}-{\delta}_{5s}\right) \left(Y_{b5}-Y_{55} \right)
\nl &&
+ ~
\left({\delta}_{bs}-{\delta}_{5s}\right) \left(Y_{a5}-Y_{55} \right)+
\frac{{s\choose 0}_5}{{0\choose 0}_5}\left(Y_{a5}-Y_{55} \right)\left(Y_{b5}-Y_{55} \right) \Bigr]
\Biggr\}I_4^{[d+],s} 
\nonumber \\ &&
+~\frac{1}{{0\choose 0}_5} \sum_{s=1}^{5} \frac{{\Sigma}^{1,s}_{b}}{{0s\choose 0s}_5} \sum_{t=1}^{5}
{\Sigma}^{2,st}_{a} I_3^{st} ,
\label{contr2}
\eea
with  $\Sigma^{1,s}_{b}$ from \eqref{eq-wa-83a} and 
\bea
\label{eq-wa-84b}
{\Sigma}^{2,st}_{a} &\equiv&
\sum_{i=1}^{4}(q_a \cdot q_i)  {0st\choose 0si}_5
\nonumber \\
&=&\frac{1}{2} \left(1-{\delta}_{st}\right)\left\{{ts\choose 0s}_5\left(Y_{a5}-Y_{55} \right) +
{0s\choose 0s}_5\left({\delta}_{at}-{\delta}_{5t}\right)-
{0s\choose 0t}_5 \left({\delta}_{as}-{\delta}_{5s} \right) \right\},
\eea
{This can be extended also to higher ranks.}
We need at most double sums, e.g.:
\bea
\label{eq-wa-82x}
{\Sigma}^{2,s}_{ab} &\equiv&
\sum_{i,j=1}^{4} (q_a \cdot q_i) (q_b \cdot q_j) {si\choose sj}_5
\nonumber\\
&=&~\frac{1}{2}(q_a \cdot q_b){s\choose s}_5
-\frac{1}{4}{\left(\right)}_5\left({\delta}_{ab}{\delta}_{as}+{\delta}_{5s} \right) .
\eea
Many of the {sums over signed minors, weighted with scalar products of chords} are given in \cite{Fleischer:2011nt}, and an almost complete list may be obtained on request from 
the authors.

\begin{figure}[t]
\begin{center}
{\includegraphics[width=.4\textwidth]{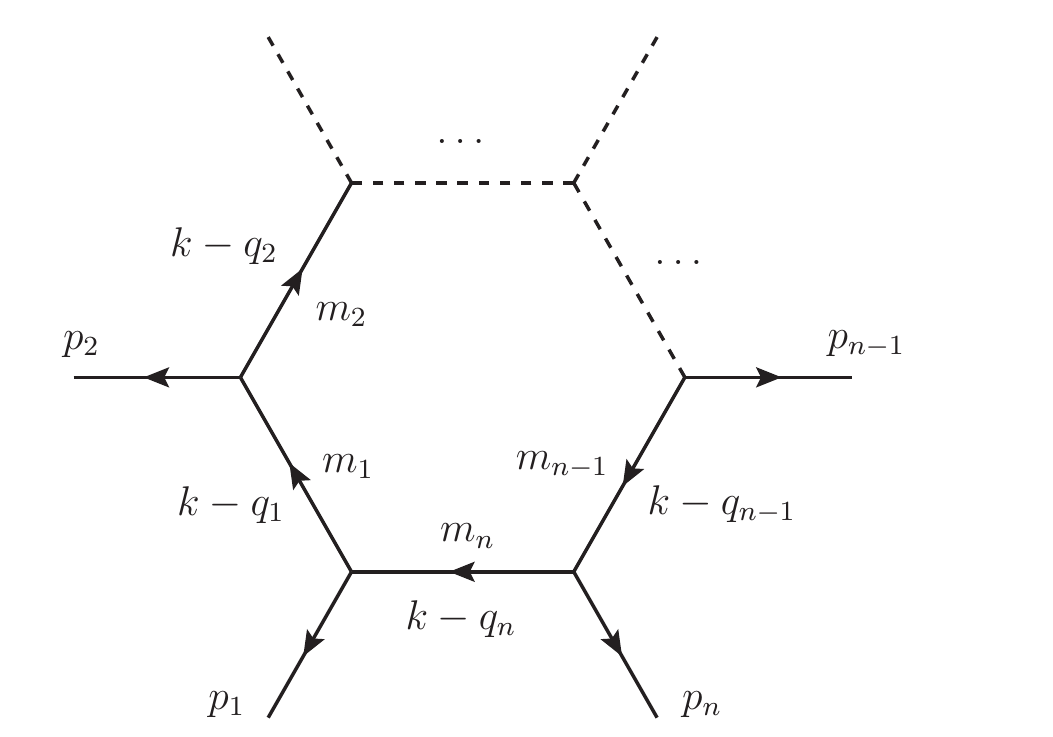}}%
\end{center}
\caption[]{%
{{Momenta definitions. }}
}
\label{momenta}
\end{figure}

\appendix
\newcommand{\snu}{s_ {\bar{\nu} u}} 
\newcommand{\ted}{t_{ed}} 
\newcommand{\tedcr}{t_{ed,\mathrm{crit}}} 
\newcommand{\smnu}{s_ {\mu \bar{\nu}   u}} 
\newcommand{\tem}{t_ {\bar{e} \mu}} 

\section{Determinants\label{app-det}}
The 
 Gram determinant $G_N$ of the diagram shown in \ref{momenta} is
\begin{eqnarray}\label{gram}
G_N &=& |2q_iq_j|, i,j=1,\ldots N .
\end{eqnarray}
 {We are also using the modified  Cayley determinant $()_N$} of a diagram with $N$ internal lines.
For the choice $q_N=0$, both determinants are related:
\begin{equation}
\label{cayl}
()_N ~\equiv~  \left|
\begin{array}{ccccc}
  0 & 1       & 1       &\ldots & 1      \\
  1 & Y_{11}  & Y_{12}  &\ldots & Y_{1N} \\
  1 & Y_{12}  & Y_{22}  &\ldots & Y_{2N} \\
  \vdots  & \vdots  & \vdots  &\ddots & \vdots \\
  1 & Y_{1N}  & Y_{2N}  &\ldots & Y_{NN}
\end{array}
\right| 
 = -G_{N-1}
  ,
\end{equation}
where  the  matrix elements are defined by
\begin{equation}
Y_{ij}=-(q_i-q_j)^2+m_i^2+m_j^2   , \quad (i,j = 1 \ldots N) .
\end{equation}
Evidently, the  {Gram} determinant $G_N$  does not depend on the masses, and so doesn't  $()_N$.

Signed minors of $()_N$ are constructed by deleting $m$ rows and $m$ columns
from $()_N$, and multiplying it with a sign factor:
\begin{eqnarray}
\label{signmino}
\left(
\begin{array}{cccc}
  j_1 & j_2 & \cdots & j_m\\
  k_1 & k_2 & \cdots & k_m\\
\end{array}
\right)_N
\equiv
{(-1)}^{\sum_l (j_l + k_l)}
\mbox{sgn}_{\{j\}} \, \mbox{sgn}_{\{k\}} \,
 \left|
\begin{array}{c}
\mbox{rows $j_1\cdots j_m$ deleted}\\
\mbox{columns $k_1\cdots k_m$ deleted}\\
\end{array}
 \right| ,
\end{eqnarray}
where $\mbox{sgn}_{\{j\}}$ and $\mbox{sgn}_{\{k\}}$ are the signs of
permutations that sort the deleted rows $j_1\cdots j_m$ and columns
$k_1\cdots k_m$ into ascending order.

Example:
\begin{eqnarray}
\label{signmino1}
\left(
\begin{array}{c}
  0 \\
  0 \\
\end{array}
\right)_N
~\equiv~
 \left|
\begin{array}{cccc}
   Y_{11}  & Y_{12}  &\ldots & Y_{1N} \\
   Y_{12}  & Y_{22}  &\ldots & Y_{2N} \\
    \vdots  & \vdots  &\ddots & \vdots \\
   Y_{1N}  & Y_{2N}  &\ldots & Y_{NN}
\end{array}
\right|   ,
\end{eqnarray}
To be definite, take  fig.~\ref{fig-4-and-6-point} as a starting point.
The example is taken from \cite{Denner:THHH2009}.
The corresponding 4-point tensor integrals are, in  LoopTools/FF \cite{Hahn:1998yk,Hahn:2010aa,vanOldenborgh:1990yc} notation:
\bea\label{n1}
\mathrm{D0i}(\mathrm{id},0,0,s_{\bar{\nu}u},t_{ed},\tem,\smnu,0,M_Z^2,0,0). 
\eea
The Gram determinant is
\bea\label{n2}
()_4
 &=&
-~2   \tem [\smnu^2 + \snu \ted - \smnu (\snu + \ted - \tem)],
 \eea
and it vanishes if:
\bea\label{n3}
\ted \to \tedcr =
\frac{ \smnu ( \smnu -  \snu +  \tem)}{ \smnu -  \snu} .
\eea
In terms of   a dimensionless scaling parameter $x$,
\bea\label{n4}
\ted &=& (1+x) \tedcr,
\eea
the Gram determinant becomes:
\bea\label{n5}
()_4 = 2 ~x~ \smnu \tem (\smnu - \snu + \tem) .
\eea
The modified Cayley determinant
\bea\label{cay1}
{0\choose 0}_4
&=&
\begin{pmatrix} 2 M_Z^2 & M_Z^2 &M_Z^2-  \smnu&M_Z^2\\ M_Z^2 &0  &-\snu& M_Z^2\\M_Z^2-  \smnu&-\snu&0&-\ted\\M_Z^2&- \tem&-\ted&0\end{pmatrix}
\nl
&=&
 \smnu^2 \tem^2 +
2~ M_Z^2 \tem [-2 \snu \ted + \smnu (\snu + \ted - \tem)]
\nonumber \\
&&+~
 M_Z^4 (\snu^2 + (\ted - \tem)^2 - 2 \snu (\ted +\tem)) 
\eea
has another dimension.
From \eqref{eqarb4} we see that  a small Gram determinant expansion will be useful when the following dimensionless parameter becomes small:
\bea\label{n6}
R &=& \frac{()_4}{{0\choose 0}_4} ~\times~ S,
\eea
where $S$ is a typical scale of the process, e.g. we will choose  $S=\smnu$.

\begin{figure}[t]
\begin{center}
{\includegraphics[width=.45\textwidth]{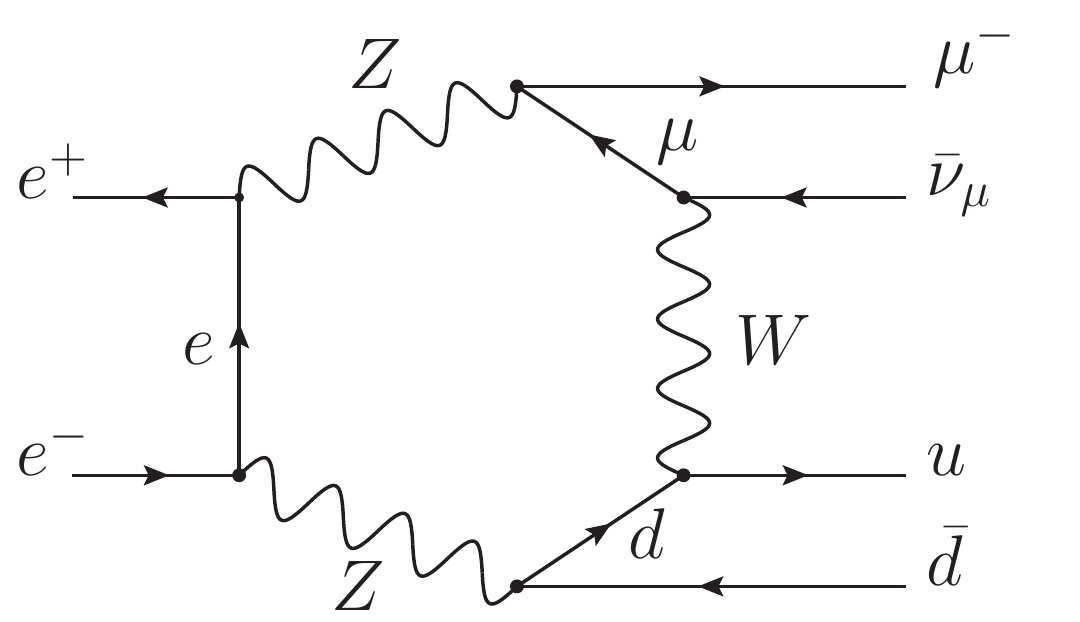}}
\hspace*{0.5cm}
{\includegraphics[width=.45\textwidth]{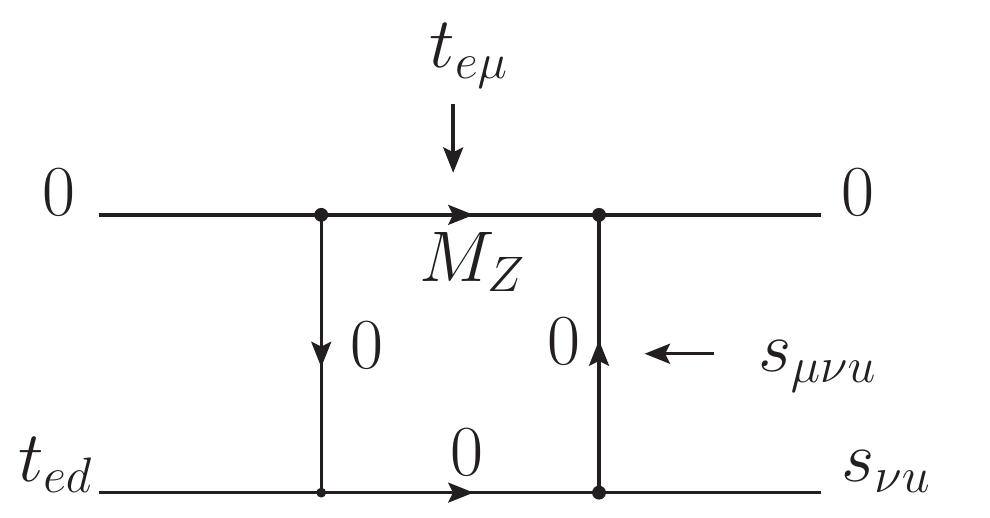}}
\end{center}
\caption[]{%
{{A six-point topology (a) leading to four-point functions (b) with realistically vanishing Gram determinants. }}
}
\label{fig-4-and-6-point}
\end{figure}

\section*{Acknowledgements}
J.F. thanks DESY for kind hospitality.
Work is supported in part by Sonderforschungsbereich/\-Trans\-re\-gio SFB/TRR 9 of DFG
``Com\-pu\-ter\-ge\-st\"utz\-te Theoretische Teil\-chen\-phy\-sik" and
European Initial Training Network LHCPHENOnet PITN-GA-2010-264564.

%


\end{document}